*Research Article*

# Social Networks through the Prism of Cognition

**Radosław Michalski**,[1] **Boleslaw K. Szymanski**,[2,3] **Przemysław Kazienko**,[1] **Christian Lebiere**,[4] **Omar Lizardo**,[5] **and Marcin Kulisiewicz**[1]

[1]*Department of Computational Intelligence, Faculty of Computer Science and Management, Wrocław University of Science and Technology, Wrocław, Poland*
[2]*Department of Computer Science, Rensselaer Polytechnic Institute, Troy, NY, USA*
[3]*Społeczna Akademia Nauk, Łódź, Poland*
[4]*Carnegie Mellon University, Pittsburgh, PA, USA*
[5]*University of California, Los Angeles, CA, USA*

Correspondence should be addressed to Radosław Michalski; radoslaw.michalski@pwr.edu.pl

Received 11 June 2020; Revised 18 November 2020; Accepted 24 December 2020; Published 8 January 2021

Academic Editor: Fei Xiong

Human relations are driven by social events—people interact, exchange information, share knowledge and emotions, and gather news from mass media. These events leave traces in human memory, the strength of which depends on cognitive factors such as emotions or attention span. Each trace continuously weakens over time unless another related event activity strengthens it. Here, we introduce a novel cognition-driven social network (CogSNet) model that accounts for cognitive aspects of social perception. The model explicitly represents each social interaction as a trace in human memory with its corresponding dynamics. The strength of the trace is the only measure of the influence that the interactions had on a person. For validation, we apply our model to NetSense data on social interactions among university students. The results show that CogSNet significantly improves the quality of modeling of human interactions in social networks.

## 1. Introduction

There is a fundamental difference between the ways social events are currently represented by network analysts and the way they are perceived and cognitively processed by humans. In the former, the discrete nature of events is retained and the weights of social network edges are updated once per each relevant event. In human memory, by contrast, perception of events changes continuously over time. Moreover, the initial strength of a trace depends on cognitive factors defined by states of mind of participants and specific aspects of their interactions. Decisions about whether to initiate, maintain, or discontinue social relations involve cognitive processes operating on all relevant information stored in memory traces over specific time-scales.

To date, however, there are no models that accurately represent the dynamics of social relations strictly through their corresponding traces in human memory. To address this gap, we introduce a novel cognition-driven social network (CogSNet) model that captures impact of human memory on perception of accumulated events and on decisions to form, maintain, or dissolve social relations. The model explicitly represents some of the human memory dynamics, such as the gradual decay of memory traces over time. With suitable data, it can be extended to include additional cognitive aspects, such as individual levels of sensitivity to relevant events, emotions, or distractions during perception of events. Hence, the model is capable of capturing the dynamics of social interactions in natural settings from the cognitive perspective of each participant.

To evaluate the empirical performance of the proposed model, we compare model's results obtained with parameters fitted using behavioral communication event data, to the ground truth perceptual data collected from human participants with regard to their most important relations. The comparisons reveal that the perception of the depth of interactions between people is well captured by the CogSNet model. At any given point in time, the model can compute the current strength of memory traces, including the impact of discrete events creating or reinforcing these traces. The computed state of each social relation (e.g., salient versus decayed) can then be compared against the externalization of cognition in the form of self-reports. The CogSNet model with parameters fitted using recorded event data achieves high accuracy in fitting ground truth data for the NetSense network. This success demonstrates the importance of incorporating cognitive processes and memory dynamics for adequate modeling of the dynamics of social relationships.

Memory is considered to be one of the most important components of human cognition. This is especially the case given the necessity to efficiently retrieve large amounts of knowledge and to select information from a noisy environment. Hence, one of the fundamental challenges for cognitive science has been to understand the mechanisms involved in managing information in human memory [1]. The exact details of these mechanisms are difficult to firmly establish since the human brain is a highly complex system featuring strong differences across individuals based on experimental tuning [2]. Yet, there have been significant advances in the direction of developing good quality working models of memory and other core cognitive processes. For example, the ACT-R cognitive architecture does a good job of modeling core features of human declarative memory, successfully replicating a large number of well-known effects. For our purposes, the most important among these empirical regularities are the well-established primacy and recency effects for list memory [3].

Limited capacity and the gradual decay of traces in memory over time have been confirmed by many studies analyzing the relationship between time and recall [4–6]. The forgetting mechanism is modeled by a parametric function describing how well a given item will be recalled as a function of time. It appears that humans tend to underestimate the number of events they experience, i.e., they actually forget faster than they think they do [5]. Obviously, any reminiscence of a particular event primed by an external situation allows people to remember the information longer. Yet ultimately, the limited capacity paradigm and the forgetting function are chiefly responsible for controlling the lifetime of such reminiscence in memory [7]. There are relatively many papers within the broader domain of link predictions, but they treat human communication as a series of fixed time widows that reflect neither human cognition nor continuity of forgetting. For example, Reader et al. [8] identified over a dozen of predictive factors quantifying phone communication like time of the last call or frequency of calls within a 4-week period that would enable to forecast persistence of social relationships, i.e., communication, also in the following 4-week period. The influences among culture, cognition, and network thinking are mentioned in [9] but in a very qualitative way, without any hint how to quantify such influence into a model, which is essential for our approach. Some other works also study the problem of recalling the names of their peers in a variety of activities [10–12], but they focus on a detailed analysis of personal attributes differentiating subjects in terms of forgetting, while here we concentrate on the models of forgetting.

## 2. Materials and Methods

The human brain records events as they arrive, but only a small fraction of the incoming information is stored in long-term memory due to its limited capacity. The forgetting mechanism dictates that the chance to recall a given event decreases gradually as we move forward from the time of first exposure [6]. In some sense, it is similar to graph-streaming [13] and feed-based social media network cascades [14] scenarios where incoming events are ordered by their arrival time, but only some of them are kept.

Accordingly, the CogSNet model uses a *forgetting function* $f$ to account for the decreasing probability of keeping aging of memory traces over time. Forgetting is thus a monotonically nonincreasing function of time with $f(0) = 1$ and $f(t) \geq 0$ for all $t > 0$. It is defined by three parameters: *reinforcement peak* $0 < \mu \leq 1$, *the forgetting intensity* $\lambda$ (both used in the ACT-R model), and additional third parameter, *forgetting threshold* $0 < \theta < \mu$.

For clarity, we present here the basic CogSNet model; a more general version is defined in Appendix and Appendix equations (A.1) and (A.2). This model processes an event happening at time $t$ as follows. If the event involves a pair of unconnected nodes $i$ and $j$, an edge $(ij)$ linking these nodes is created. This edge is assigned weight $w_{ij}(t)$ equal to the reinforcement peak value $\mu$. Otherwise, if the nodes involved are already connected, the forgetting function $f$ is used to set the weight of the edge connecting these nodes:

$$w_{ij}(t) = \begin{cases} \mu, & \text{if } w_{ij}(t_{ij}) f(t - t_{ij}) < \theta, \\ \mu + w_{ij}(t_{ij}) f(t - t_{ij})(1 - \mu), & \text{otherwise}, \end{cases} \quad (1)$$

where $t_{ij}$ is the time of the preceding event for this edge. Processing of each event ends with advancing $t_{ij}$ to $t$. Initially, $t_{ij}$ and $w_{ij}(0)$ are set to 0. Equation (1) and limit on value of $\mu$ ensure that a weight of any edge is at most 1.

The weight $w_{ij}(t)$ of an edge $(ij)$ between two nodes at any user selected time $t$ is computed as follows. Once all the relevant events up to time $t$ are processed, we simply set $w_{ij}(t) = w_{ij}(t_{ij}) * f(t - t_{ij})$. If the result is less than the forgetting threshold $\theta$, $w_{ij}(t)$ is reduced to zero and the edge is removed. A threshold is needed with forgetting functions, such as power and exponential forgetting, that are positive for nonnegative arguments. Otherwise, an edge would get positive weight at creation and would always stay positive; i.e., all created memory traces would never cease to exist. The reinforcement peak $\mu$ defines the impact of an event on the weight of the edge relevant to this event. This peak is a global model parameter here. In principle, the peak can be adjusted according to the event or node type to allow for individualized event perception.

In Figure 1, we compare the CogSNet model with the previous proposals for representing temporal network dynamics. The most common approach for representing social network dynamics is to use interaction sequences [15]. Under this method, each event is time-stamped and the weights are added to the edges connecting nodes involved in this event cf. Figure 1 ((I) and (IIa)). Moreover, a given edge is active (exists) only at a given time $t$. This is the most granular approach as it is capable of tracking all the events occurring between nodes while preserving the temporal order of events.

In contrast, a static binary network representation, as shown in Figure 1 (IIb), aggregates all events by making all edges time-independent. Consequently, an edge existing between a pair of nodes corresponds to an event between these nodes occurred at least once in the whole observed period [16]. Such an edge representation throws away information on the temporal ordering of events, making it impossible to study dynamic processes in static networks.

The incremental network solution accumulates events only up to the current time $t$ of analysis. The classical approach, used early in [17–19], views a dynamic network as a series of time-ordered sequences of static graphs (see Figure 1 (IIc)). More recently, this method was applied to modeling network and community evolution [20, 21]. The drawback of this approach is that it does not preserve the ordering of interactions within time slices. Applying a simple frequency-based aggregation creates a frequency-based (FQ) metric cf. Figure 1 (IId). Taking into account only a given number of the most recent events leads to the recency-based (RC) model cf. Figure 1 (IIe). Both of these models are used here as baseline models.

Figure 1 (IIf) shows an example of dynamic social network generated using the CogSNet model. All other social network models presented in Figure 1 ((IIa)–(IIe)) can also be represented by CogSNet by setting appropriate parameter combinations to achieve, as needed, no decay, instant decay, and so forth. In this way, CogSNet can be thought of as a universal generative dynamic model for temporal social networks, encompassing previous approaches as special cases.

In general, the forgetting function $f(\Delta t)$ over time interval $\Delta t$ can be of any type (linear, power, and logarithmic), but here, informed by work in the cognitive psychology of memory [5], we evaluate only two such functions: the exponential function $f^{\exp}$ and the power function $f^{\mathrm{pow}}$ defined as follows:

$$f^{\exp}(\Delta t) = e^{-\lambda \Delta t}, \quad (2)$$

$$f^{\mathrm{pow}}(\Delta t) = \max(1, \Delta t)^{-\lambda}, \quad (3)$$

where $\lambda$ denotes the forgetting intensity; typically, $\lambda \in [0, 1]$. The use of max in the power function ensures that perception of events that happened less than a time unit ago is not changed by forgetting. The time unit in which the forgetting function is expressed scales the values of the parameters. Our experiments use one hour as the time unit.

To simplify optimal parameters search, we aggregate all three parameters into the *trace lifetime* $L$ defining the time after which an unreinforced memory trace is forgotten, i.e., is too weak to be recalled. In the model, $L$ is the time over which the forgetting function reduces the edge weight from $\mu$ to $\theta$ causing the edge to be removed cf. Figure 2. For the exponential forgetting function (equation (2)), trace lifetime $L^{\exp}$ is

$$L^{\exp} = \frac{1}{\lambda} \ln\left(\frac{\mu}{\theta}\right), \quad (4)$$

while for the power function, the formula is

$$L^{\mathrm{pow}} = \left(\frac{\mu}{\theta}\right)^{1/\lambda}. \quad (5)$$

Please note that according to the CogSNet model the relationship weights belong to range $[\theta; 1) \cup \{0\}$. Edge's weight value tends to 1 when the node encounters large number of tightly successive contacts. The larger the number of events is and the closer to each other they are, the closer to 1 the weight is.

## 3. Results and Discussion

In this section, we assess the empirical validity of the propose model by estimating the accuracy with which the CogSNet model reproduces the dynamics of social relations in a dynamic social network. Our computational experiments use the NetSense dataset introduced in [22].

The data contain 7,096,844 human mobile phone communication events, including both calls and text messages. These are augmented by 578 surveys containing self-reports on top contacts. It is worth underlining that this dataset provides a unique opportunity to evaluate how the communication frequencies are recalled by the study participants because students reported their perception of the frequency—detailed information on how this information has been reported and how data were processed is provided in Appendix.

We use this dataset to study the evolution of two coupled social networks of university students. The first is a behavioral network representing interactions between individuals in the form of the records of their mobile calls and text messages. The second one has perceptual edges defined by the personal nominations. These nominations are based on students' perception of the corresponding relations as one of the top twenty most interacting peers in the surveys administered to participants. These surveys cover the first four semesters of the student's college experience (beginning of freshman year to the end of the sophomore year). We compare the list of nominations predicted from the CogSNet network model purely from the communication event data with the list of nominations collected in a given survey.

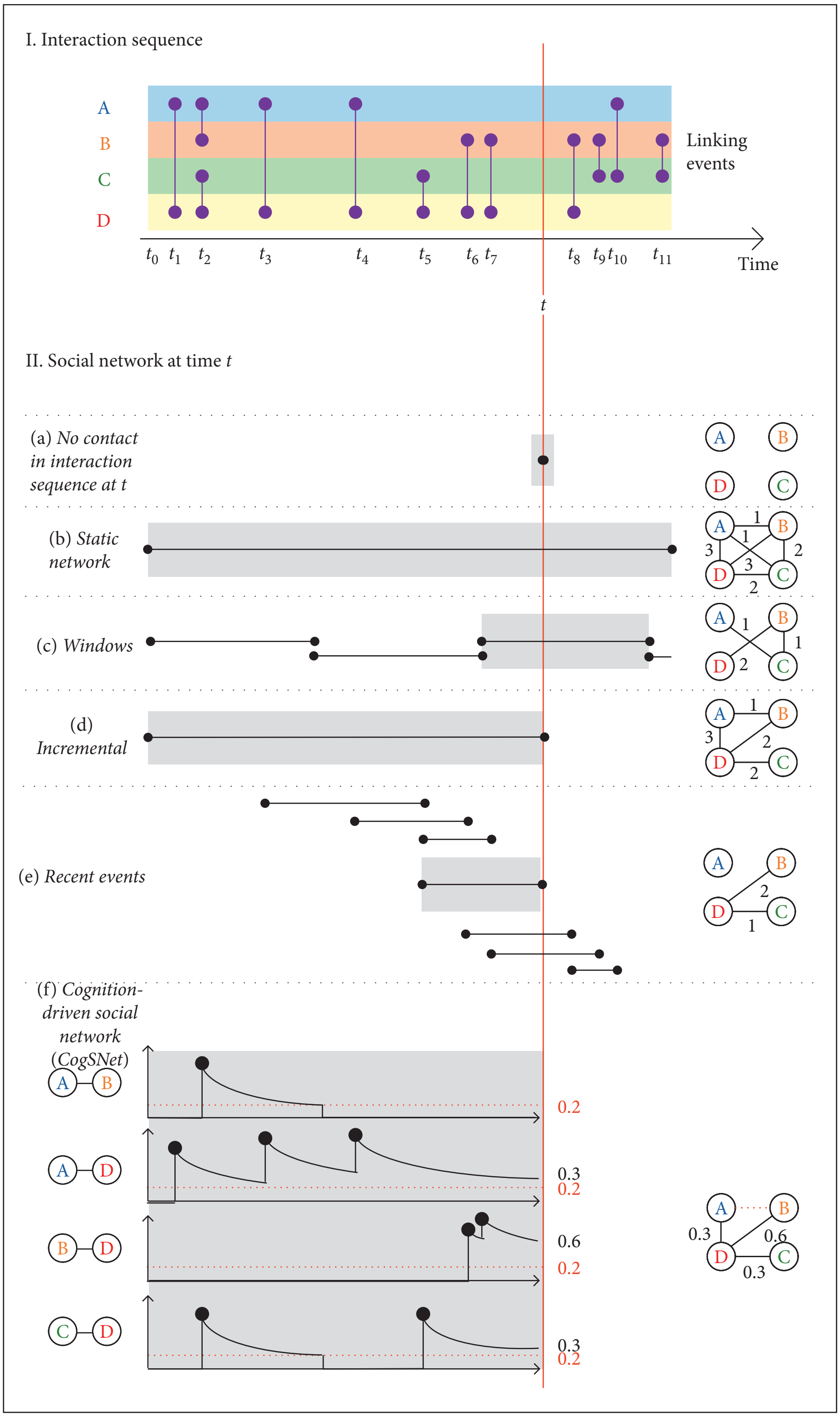

FIGURE 1: Various approaches to modeling dynamic networks and edge weighting (relation strengths) for the 4-node network at a given time $t$: (I) and (IIa) interaction sequence; (IIb) static (time-aggregated) network; (IIc) sliding windows; (IId) incremental network, all events from the beginning time $t_0$ to the current time $t$ are considered, and the frequency of interaction in the period constitutes the frequency-based (FQ) reference; (IIe) network based on $n = 3$ recent events, used for recency-based reference RC; (IIf) cognition-based social network model (CogSNet), introduced here.

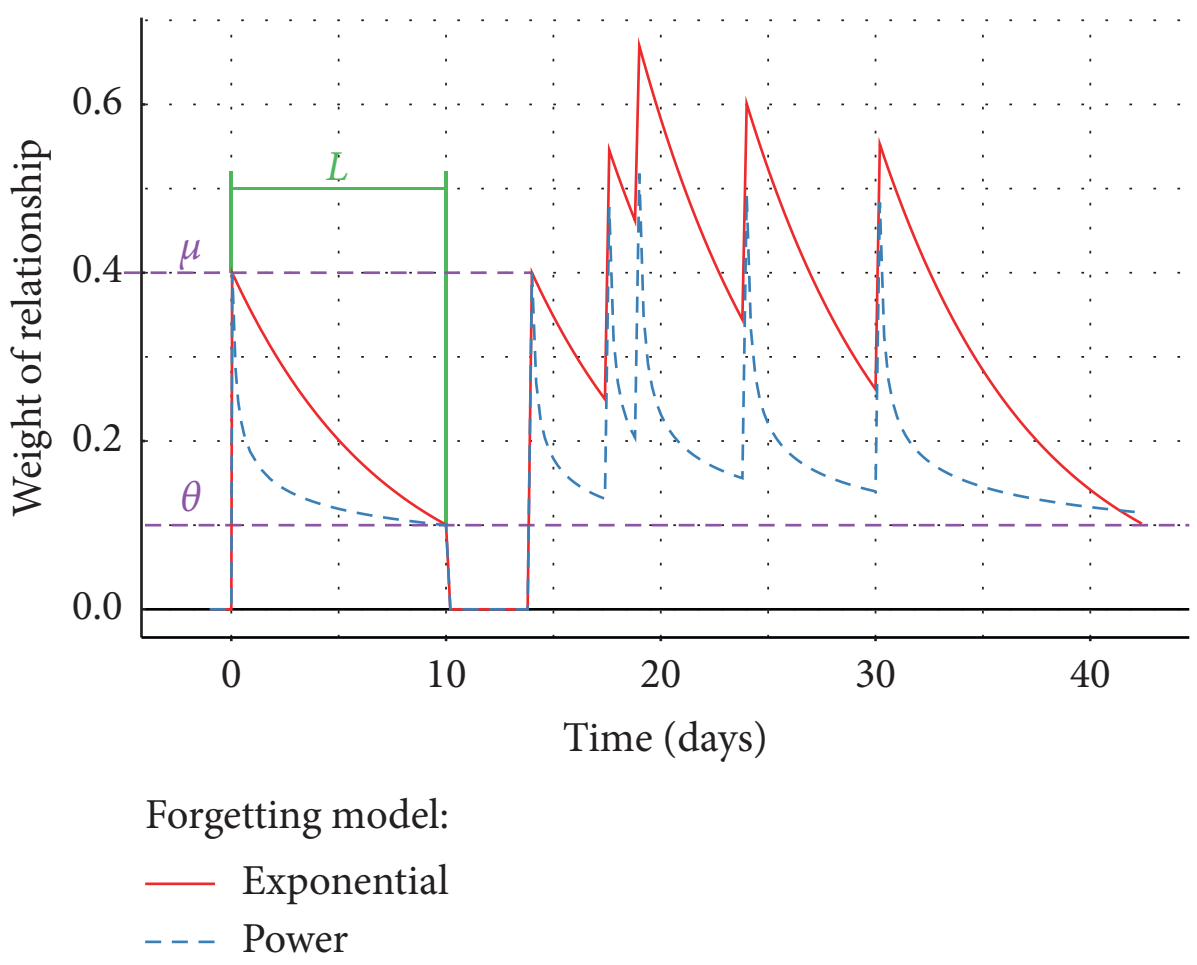

Figure 2: Dynamics of relations in CogSNet network with exponential and power functions and with parameters set to $\mu = 0.4$, $\theta = 0.1$, and $L = 10$ days.

Figure 3 shows an example of dynamic social network generated from a subset of NetSense data using the CogSNet model.

For comparison, we also implement three baseline models using the mobile phone communication data. The first is a *frequency-based* (FQ) model which orders the peers by the number of interactions, regardless of their time order. Under this model, the highest ranking peers whose number equals the number of the nominations listed in the corresponding survey are selected on the basis of their communication frequency. The *recency-based* (RC) model selects the top interacting peers who for each individual given a number of recent events. Finally, the *random sampling* (RND) model creates the list by randomly selecting peers from those who are recorded in the history of interactions of a given node. While random baseline can be considered as a simplistic lower bound, the two aforementioned are actually regarded to be state of the art since they express two cognitive phenomena. The recency effect was uncovered by the experiments of Hermann Ebbinghaus on recall. The forgetting curve he proposed is based on studies in [23] and recently confirmed in [24]. In these papers, the authors demonstrate that the learning curve rises the chances for recalling at the beginning and at the end of the contact period, while the chances to recall events are lower in the middle of it. However, as the question in surveys mainly refers to the present, we focused on the recency instead of primacy. The second baseline, frequency-based, is an implementation of scientific results taken from the domain of lexical memory [25]. The authors of this study show that the frequently repeated words are being recalled faster. Unfortunately, to the best of our knowledge, there are no other approaches suitable to be used in similar tasks.

To compare the performance of all the models, we use a *Jaccard* metric, equation (C.1), in Appendix, which measures the ratio of the number of nominations produced by the model that are also ground truth nominations listed in the corresponding survey divided by the number of unique nominations on both lists. Due to the nature of the question formulation in the survey, it was impossible to use any approach based on ordering such as Kendall's or Pearson's ranks since we had to consider the list as unordered.

Figure 4 shows the results of this comparison over the range of parameters corresponding to reported values for forgetting of one to 14 weeks. As reported in [26], the ability to recall information about social interactions starts to degrade after about one week. The experiments using NetSense dataset reveal that the performance is the highest when the forgetting of unreinforced memory traces happens after two weeks. The results remain satisfactory for forgetting thresholds as low as one week. With the two-week threshold, the CogSNet model is with either power or exponential forgetting statistically significantly better than any baseline model. In the literature, the power forgetting was found to perform significantly better than the exponential forgetting when a range of parameters was limited [3]. We do not observe such superior performance of power forgetting here. Both functions have similar peak of Jaccard metric, albeit for the different lifetime values.

When comparing the results of surveys with the states of the CogSNet network at the times of the surveys, the Jaccard metric is as high as 29.98% for exponential function for $L = 3$ days. The distant second is recency-based RC model which delivers much lower Jaccard metric of 17.8%. Nevertheless, when looking at relative simplicity of the RC metric, one can also say that it performs quite good—this is due to the fact

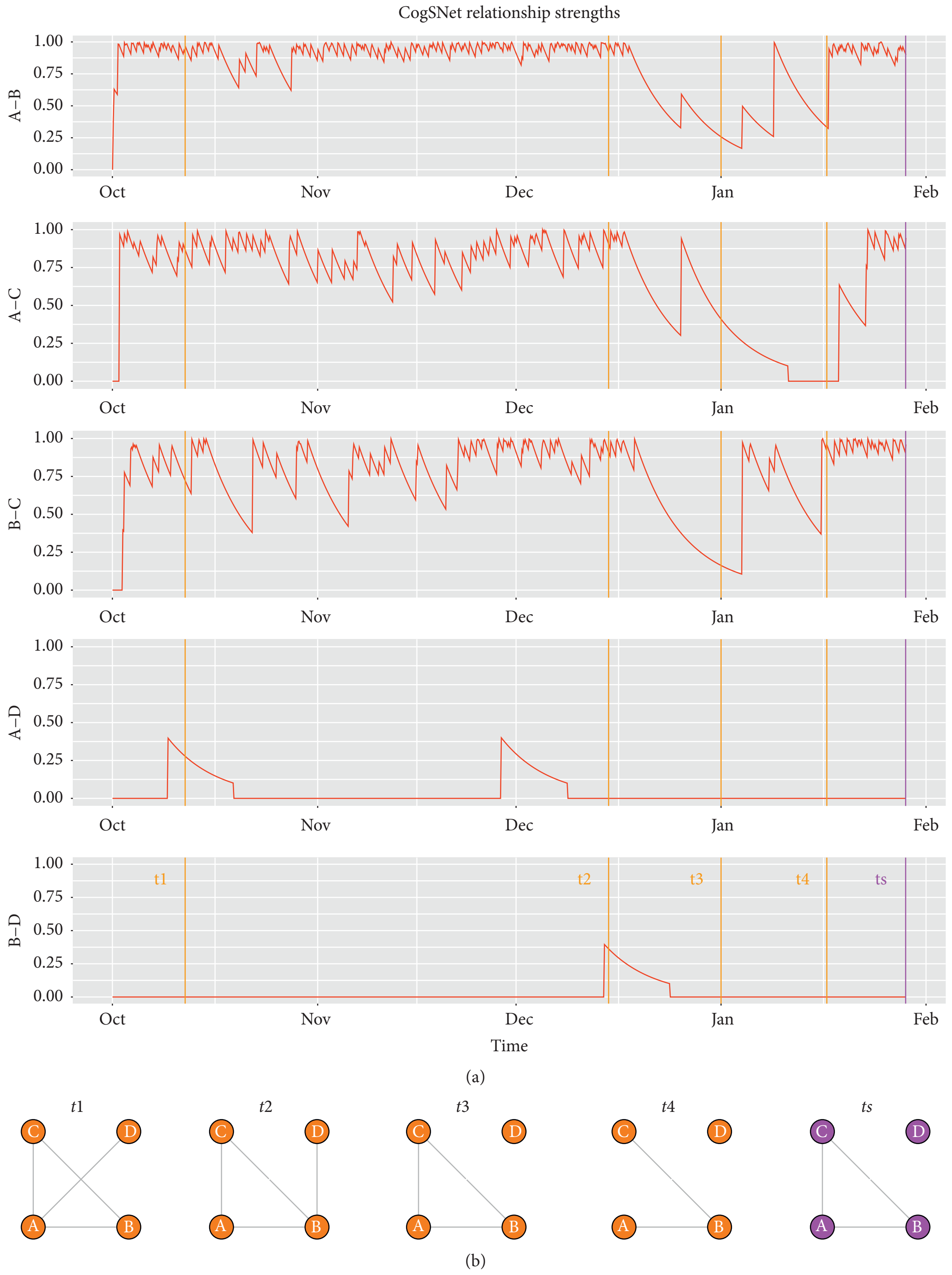

Figure 3: Four-node CogSNet network for the sample of real NetSense data: $\mu = 0.4$, $\theta = 0.1$, and $L = 10$ days; nodes A, B, C, and D correspond, respectively, to participants with ids 40997, 11360, 10841, and 1232. (a) Relation strengths according to the CogSNet model over 4-month period (one term); (b) network snapshots at four timestamps $t_1$–$t_4$ and at the survey time $t_s$.

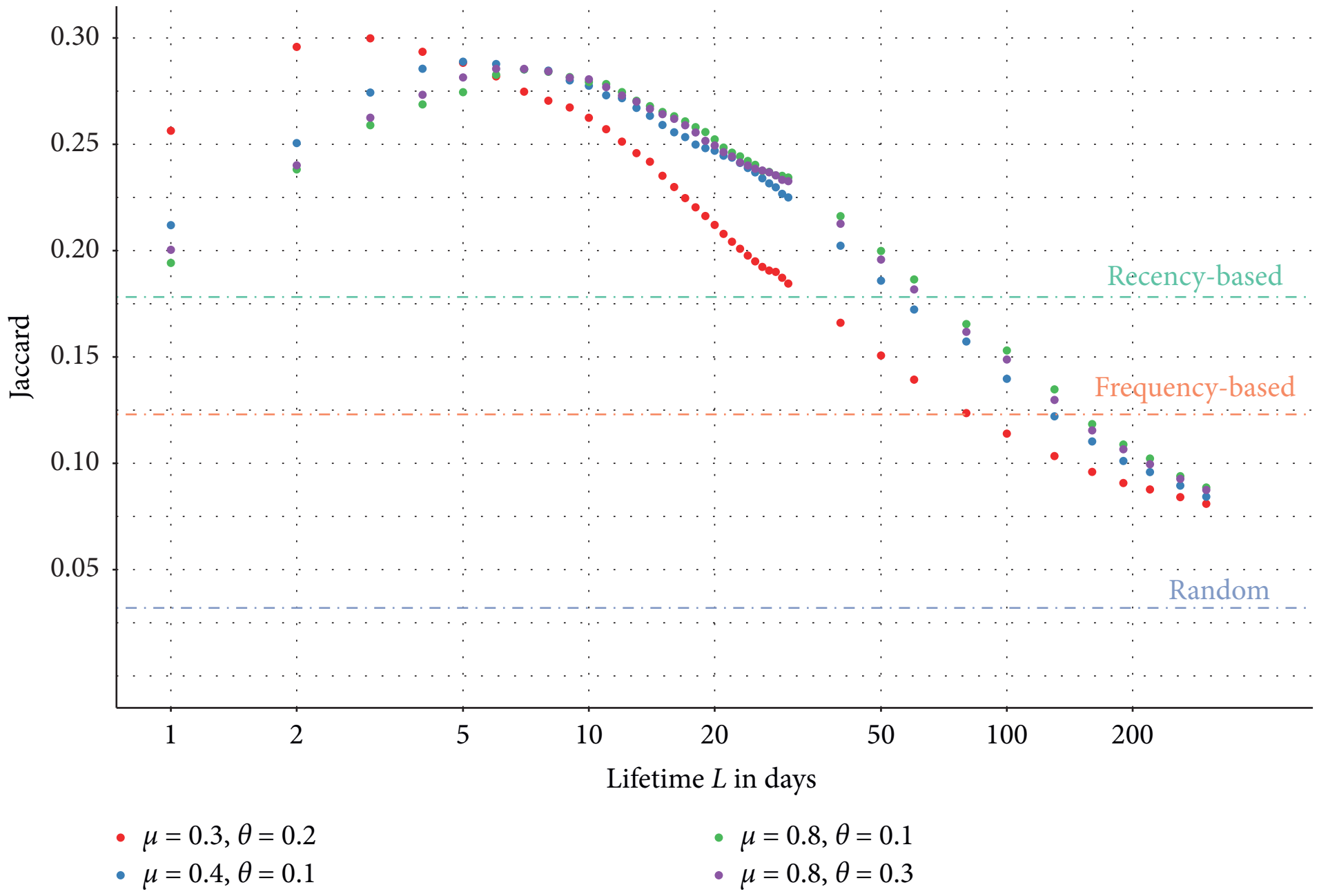

(a)

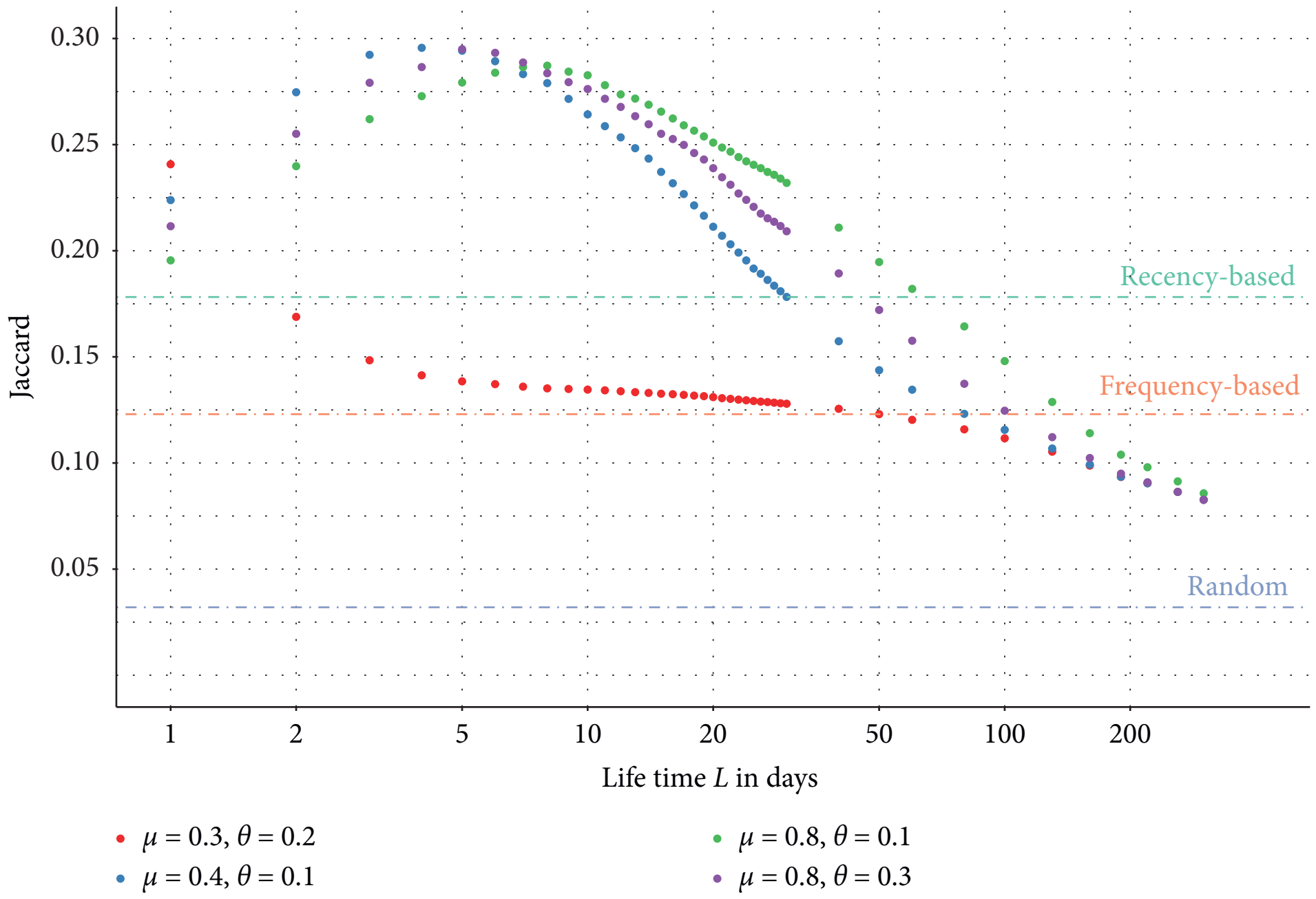

(b)

Figure 4: The plot of the Jaccard metric as a function of lifetime $L$. The metric measures the overlap between the two sets of peers, one identified by CogSNet and the other listed by students in the survey. These plots are compared to the results achieved by the three baseline models: recency-based with the best results obtained with the number of recent events set to 400, frequency-based, and random. The results are plotted for the CogSNet running with various parameters for (a) exponential and (b) power forgetting functions.

Table 1: Numerical and symbolic $p$ values adjusted by Simes–Hochberg step-up method of nonparametric pairwise post hoc Nemenyi test comparing values of Jaccard metric for results obtained by each method with parameters fitted using recorded event data with the ground truth based on surveys taken over semesters 1–4.

| | Power | Exponential | Recency | Frequency | Random |
|---|---|---|---|---|---|
| Power | — | *** | *** | *** | *** |
| Exponential | 0.0000 | — | *** | *** | *** |
| Recency | 0.0000 | 0.0000 | — | *** | *** |
| Frequency | 0.0000 | 0.0000 | 0.0000 | — | *** |
| Random | 0.0000 | 0.0000 | 0.0000 | 0.0000 | — |

These results show how each method in the leftmost column performs versus the method in the first row. The part of table under the diagonal shows numerical values of the test, while the part above the diagonal presents $p$ values coded as follows: $^{***}p < 0.00005$ represents high confidence in the performance differences between the methods since typically values of $p > 0.05$ are considered indicative of nonsignificance.

Table 2: Setting of parameters for power forgetting function in computational experiments.

| $\lambda$ dependence on $\mu, \theta$, and $L$ for power forgetting | | | |
|---|---|---|---|
| Lifetime $L$ days | $\mu = 0.4,\ \theta = 0.1$ | $\mu = 0.8,\ \theta = 0.3$ | $\mu = 0.8,\ \theta = 0.1$ |
| 1 | 0.43621 | 0.30863 | 0.65431 |
| 2 | 0.3581 | 0.25337 | 0.53716 |
| 3 | 0.32415 | 0.22934 | 0.48623 |
| 4 | 0.30372 | 0.21489 | 0.45558 |
| 5 | 0.28957 | 0.20487 | 0.43435 |
| 6 | 0.27894 | 0.19736 | 0.41841 |
| 7 | 0.27055 | 0.19142 | 0.40583 |
| 8 | 0.26368 | 0.18656 | 0.39552 |
| 9 | 0.2579 | 0.18247 | 0.38685 |
| 10 | 0.25294 | 0.17896 | 0.37942 |
| 11 | 0.24862 | 0.1759 | 0.37293 |
| 12 | 0.2448 | 0.1732 | 0.3672 |
| 13 | 0.24139 | 0.17079 | 0.36208 |
| 14 | 0.23831 | 0.16861 | 0.35747 |
| 15 | 0.23552 | 0.16663 | 0.35328 |
| 16 | 0.23297 | 0.16483 | 0.34945 |
| 17 | 0.23062 | 0.16317 | 0.34592 |
| 18 | 0.22844 | 0.16163 | 0.34267 |
| 19 | 0.22643 | 0.1602 | 0.33964 |
| 20 | 0.22455 | 0.15887 | 0.33682 |
| 21 | 0.22278 | 0.15762 | 0.33418 |
| 22 | 0.22113 | 0.15645 | 0.3317 |
| 23 | 0.21957 | 0.15535 | 0.32936 |
| 24 | 0.2181 | 0.15431 | 0.32716 |
| 25 | 0.21671 | 0.15333 | 0.32507 |
| 26 | 0.21539 | 0.15239 | 0.32309 |
| 27 | 0.21414 | 0.15151 | 0.3212 |
| 28 | 0.21294 | 0.15066 | 0.31941 |
| 29 | 0.2118 | 0.14985 | 0.3177 |
| 30 | 0.21071 | 0.14908 | 0.31606 |
| 40 | 0.20188 | 0.14283 | 0.30282 |
| 50 | 0.19553 | 0.13834 | 0.29329 |
| 60 | 0.19062 | 0.13487 | 0.28594 |
| 80 | 0.18337 | 0.12974 | 0.27506 |
| 100 | 0.17811 | 0.12602 | 0.26717 |

Table 3: Setting of parameters for exponential forgetting function in computational experiments.

| $\lambda$ dependence on $\mu, \theta$, and $L$ for exponential forgetting | | | |
|---|---|---|---|
| Lifetime $L$ (days) | $\mu = 0.4$, $\theta = 0.1$ | $\mu = 0.8$, $\theta = 0.3$ | $\mu = 0.8$, $\theta = 0.1$ |
| 1 | 0.05776 | 0.04087 | 0.08664 |
| 2 | 0.02888 | 0.02043 | 0.04332 |
| 3 | 0.01925 | 0.01362 | 0.02888 |
| 4 | 0.01444 | 0.01022 | 0.02166 |
| 5 | 0.01155 | 0.00817 | 0.01733 |
| 6 | 0.00963 | 0.00681 | 0.01444 |
| 7 | 0.00825 | 0.00584 | 0.01238 |
| 8 | 0.00722 | 0.00511 | 0.01083 |
| 9 | 0.00642 | 0.00454 | 0.00963 |
| 10 | 0.00578 | 0.00409 | 0.00866 |
| 11 | 0.00525 | 0.00372 | 0.00788 |
| 12 | 0.00481 | 0.00341 | 0.00722 |
| 13 | 0.00444 | 0.00314 | 0.00666 |
| 14 | 0.00413 | 0.00292 | 0.00619 |
| 15 | 0.00385 | 0.00272 | 0.00578 |
| 16 | 0.00361 | 0.00255 | 0.00542 |
| 17 | 0.0034 | 0.0024 | 0.0051 |
| 18 | 0.00321 | 0.00227 | 0.00481 |
| 19 | 0.00304 | 0.00215 | 0.00456 |
| 20 | 0.00289 | 0.00204 | 0.00433 |
| 21 | 0.00275 | 0.00195 | 0.00413 |
| 22 | 0.00263 | 0.00186 | 0.00394 |
| 23 | 0.00251 | 0.00178 | 0.00377 |
| 24 | 0.00241 | 0.0017 | 0.00361 |
| 25 | 0.00231 | 0.00163 | 0.00347 |
| 26 | 0.00222 | 0.00157 | 0.00333 |
| 27 | 0.00214 | 0.00151 | 0.00321 |
| 28 | 0.00206 | 0.00146 | 0.00309 |
| 29 | 0.00199 | 0.00141 | 0.00299 |
| 30 | 0.00193 | 0.00136 | 0.00289 |
| 40 | 0.00144 | 0.00102 | 0.00217 |
| 50 | 0.00116 | 0.00082 | 0.00173 |
| 60 | 0.00096 | 0.00068 | 0.00144 |
| 80 | 0.00072 | 0.00051 | 0.00108 |
| 100 | 0.00058 | 0.00041 | 0.00087 |

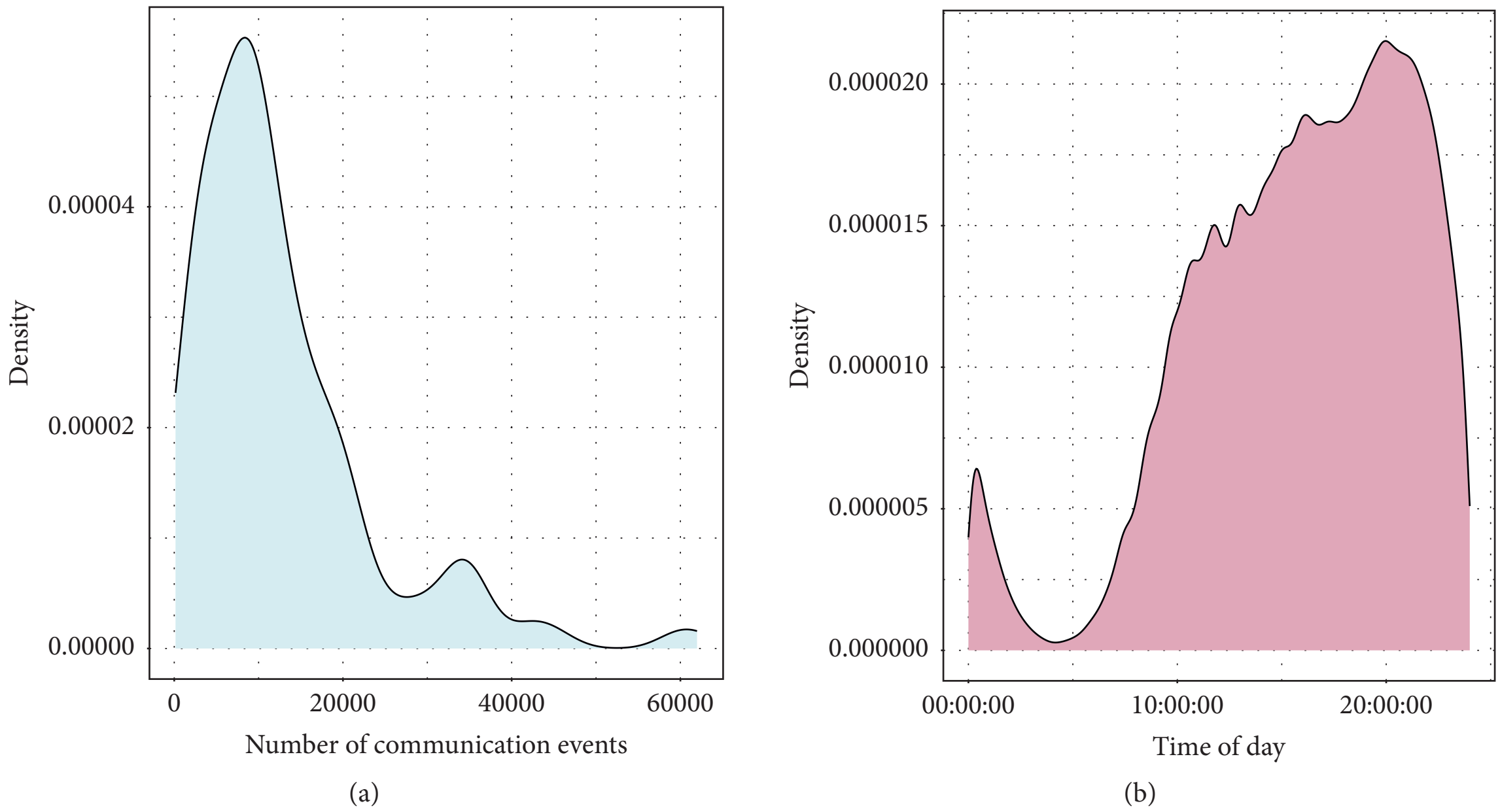

Figure 5: Continued.

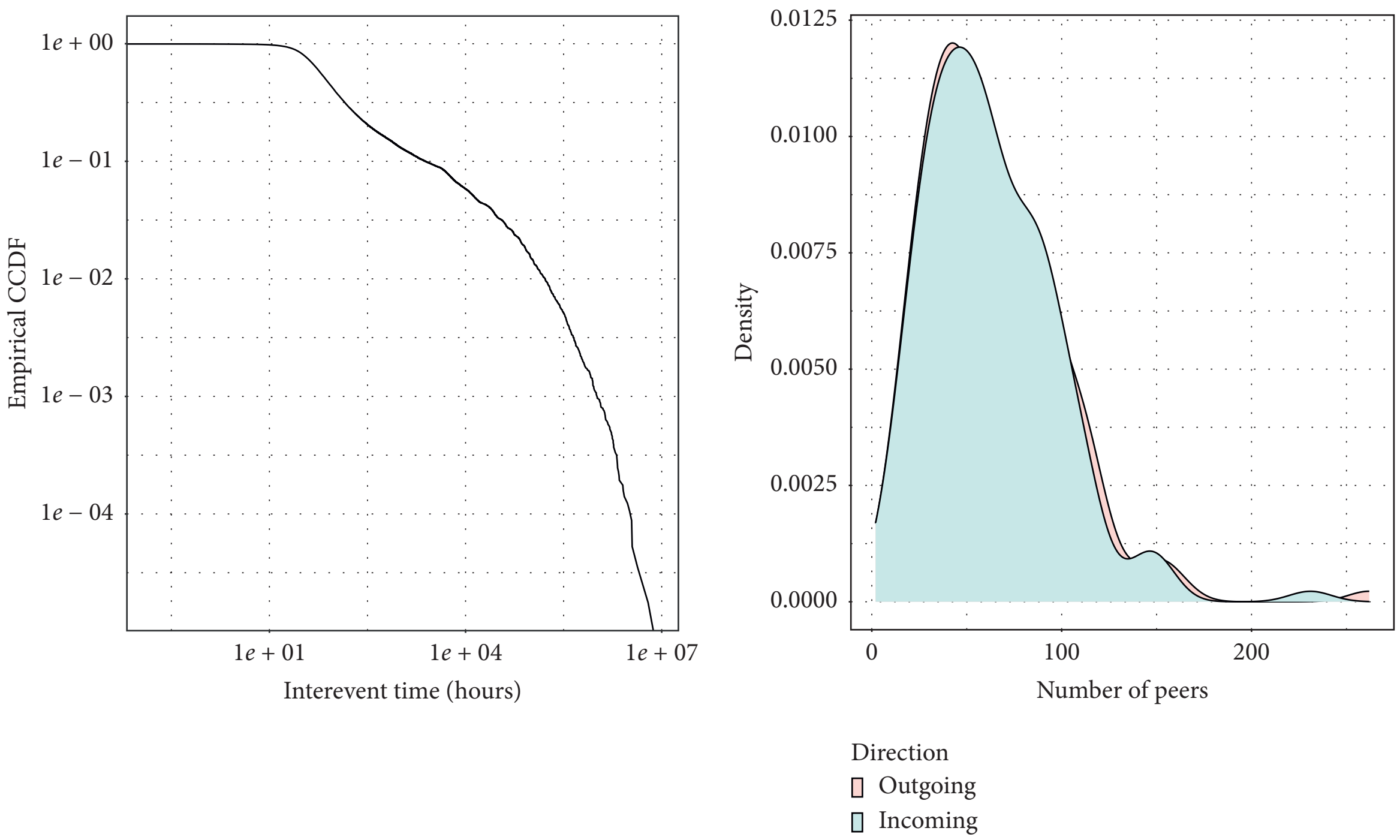

(c) (d)

Figure 5: The exploratory analysis of the communication data for students participating in the NetSense study at the University of Notre Dame: (a) density plot for the number of communication events for the students; (b) the distribution of phone activity through the course of the day; (c) the cumulative distribution function of the interevent times in hours; (d) the number of incoming and outgoing communication peers for the students in NetSense, where there are at least twenty communication events with the peer during the analyzed period [30].

that primacy and recency effects come into play when listing peers [27].

## 4. Conclusions

The results demonstrate that accounting for event perception and memory dynamics is essential for faithfully modeling the interaction dynamics in social networks. The CogSNet performed statistically significantly better than any of the base temporal models (see Table 1 in Appendix). With suitable data, the model can be extended to represent individual sensitivity to events and the impact of emotions and distractions on event perception to further improve modeling quality.

The dataset on which we operate contains data limited to a relatively small fraction of human social activities. Hence, it cannot be stated that there were no other events happening that could have affected participant memory about others. Examples of such events are face to face interactions, indirect references to someone else, and personal reminiscences about others. In addition, emotions can also play a significant role in these processes, and they can be potentially identified with some confidence from available data [28].

As a result, the reinforcement peak value can be adjusted for a given person and for an individual event cf. also equation (A.1) in Appendix. What is observed here is a partial manifestation from human memory. However, even taking into account that the model was built based on a single data source, over 6 million telephone calls and messages among the NetSense study participants, the model fitted with the recorded event data yields results well matching the salience of social contacts over all 578 surveys completed by 184 participants. It is observed that even when using the same values of model parameters for all participants, the results are satisfactory. Nevertheless, this accuracy most likely could have been increased if the parameters had been individually adjusted for each participant. Apart from that, one should remember that the recall task is highly contextual, so the individuals will provide different sets of peers depending on the situation [11, 12].

The findings of this work can be also considered as an important building block on how to maintain healthy relationships or how to discover anomalies. For instance, based on the results, it is anticipated that, in order to keep a living relationship with a peer, the contacts should not be sparser than lifetime $L$, preferably even more frequent. On the contrary, these findings also demonstrate on how long a grieving period after breaking the relationship could potentially last and how to handle the recovery process afterwards in order to make it effective. Lastly, small modifications making the model closer to ACT-R would enable its use for instance in the case of Alzheimer disease where one could monitor whether the peers recalled by patients are in line with what a CogSNet model generates based on the contacts or visits. If there will be any discrepancy found between the model and observations based on interview, this could be an early sign or disease emerging or becoming more active. Another interesting case would be to see whether the strengths of new links are decaying slower than the old ones.

In future work, we plan to extend the model in the directions outlined in Appendix including accounting for distractions during interactions, individualized strength, asymmetric of interactions of significance to participants (e.g., hierarchical relationships), and the impact of forms of interactions and of associated emotions. Hence, the CogSNet model represents an important first step towards modeling social network dynamics through the prism of human cognition. In literature, there are three approaches that are used widely to understand human cognition: cognitive bandwidth (CB), dual-process morality (DPM), and implicit association tests (IAT). In [29], those methods are compared and their common drawback is pointed out—the lack of perspective that takes a cultural background into account. Our approach makes such considerations possible, and we see this challenge as one of the future work directions. Lastly, we would love to see more datasets that are so extensive as the NetSense dataset that could allow us to evaluate our model.

# Appendix

## (A) Formal Definition of the CogSNet Model

A social network can be represented by a graph $G = (V, E)$, where $V = \{v_1, \ldots, v_n\}$ denotes the set of $n$ nodes and $E = \{e_1, \ldots, e_m\}$ is the set of directed $m$ edges between pairs of nodes. Each edge $e_{ij}$ from node $v_i$ to node $v_j$, $i \neq j$, is assigned weight $w_{ij}(t)$ and represented by a triple $(v_i, v_j, w_{ij}(t))$.

The model evolves in discrete steps as follows. For each pair of nodes $(v_i, v_j)$, the system maintains two variables: $t_{ij}$, which represents the time of the most recent event for this pair of nodes and $c_{ij}$ which holds the count of events processed for this pair of nodes. Initially, both $t_{ij}$ and $c_{ij}$ are set to 0, as are the weights of all edges, i.e., for all pairs of nodes $(v_i, v_j)$, $w_{ij}(0) = 0$.

When an event happens at time $t$ in the modeled social network, it is processed in chronological order by the model. First, the weight of the corresponding edge is updated according to the following equation:

$$w_{ij}(t) = \begin{cases} \mu_{ijc_{ij}+1}, & \text{if } w_{ij}(t_{ij}) f(t - t_{ij}) < \theta, \\ \mu_{ijc_{ij}+1} + w_{ij}(t_{ij}) f(t - t_{ij}) \left(1 - \mu_{ijc_{ij}+1}\right), & \text{otherwise,} \end{cases} \quad \text{(A.1)}$$

where $\mu_{ijk}$ is the value of reinforcement peak that results from the $k^{\text{th}}$ event that impacts the edge $(v_i, v_j)$. Here, the value of reinforcement peak $\mu_{ijk}$ depends on the engagement and emotions invoked by the event that is either directly or indirectly related to the edge $(v_i, v_j)$. An example of an event indirectly related to this edge could be node $v_i$ talking about node $v_j$ or any situation that reminds node $v_i$ about node $v_j$. The values of $\mu$ can be individualized to node $v_i$ perception of relation with node $v_j$ at event $k$. The $\mu$'s values may also be dependent on event types: $\mu_{ijk} \in \{\mu_1, \mu_2, \mu_3, \mu_4, \ldots\}$, e.g., $\mu_1 = 0.5$ for emails, $\mu_2 = 0.55$ for phone calls, $\mu_3 = 0.8$ for meetings, and $\mu_4 = 0.9$ for joint collaboration in projects.

Finally, the processing of the current event updates both variables associated with the updated edge $(v_i, v_j)$ as follows: $t_{ij} = t, c_{ij} = c_{ij} + 1$.

At any time $t$ of the model evolution, the user can obtain the value of the weight of an arbitrary edge $(v_i, v_j)$ by computing the following equation:

$$w_{ij}(t) = \begin{cases} 0, & \text{if } w_{ij}(t_{ij}) f(t - t_{ij}) < \theta, \\ w_{ij}(t_{ij}) f(t - t_{ij}), & \text{otherwise.} \end{cases} \quad \text{(A.2)}$$

## (B) Experimental Parameter Space

To analyze the CogSNet model and compare its results to those obtained with the reference models, many combinations of the CogSNet parameters were tested. First, different values of $\mu$, $\theta$, and $L$ were specified, and then, $\lambda$ coefficient was calculated using equations (4) and (5), which were also used to define CogSNet parameters for experimental verification. The complete list of parameters used is listed in Tables 2 and 3.

## (C) Quality Measures

The values of Jaccard metric for a single surveyed student participant $v_i$ have been computed as follows:

$$\text{Jaccard}(v_i) = \frac{\left|V_i^{\text{CogSNet}} \cap V_i^{\text{survey}}\right|}{\left|V_i^{\text{CogSNet}} \cup V_i^{\text{survey}}\right|}, \quad \text{(C.1)}$$

where $V_i^{\text{survey}}$ is the set of up to 20 peers enumerated in the survey by the participating student $v_i$ and $V_i^{\text{CogSNet}}$ is the set of neighbors of this student in the CogSNet network with the largest nonzero weights on edges to this student on the day on which the given survey was administered.

### (D) Recency-Based Baseline

The recency-based baseline model was tuned by finding the best parameter representing the number of the most recent events to be taken into account in computing the performance. The resulting value of 400 events was used in Figure 4 for computing the results for the recency-based baseline model.

### (E) Statistical Tests

To confirm that there is a statistical difference between the proposed CogSNet model and baselines, we performed a number of statistical tests. The results are presented in Table 1 which shows that CogSNet approach is statistically significantly better from all baseline methods. Moreover, power forgetting performs statistically significantly better than exponential forgetting.

### (F) NetSense Dataset

The dataset consists of two parts. The first includes the timestamps and duration/length of phone calls and text messages collected for each student participating in the study. Each student phone device recorded all connections/messages, including those to the phones of people outside of the test group, so recording was done on both sides of communication, by sending and receiving calls/messages, if both belonged to the study. In total, the mobile phone communication data consist of 7,575,865 activities. Out of these, 7,096,844 have been text messages (93.7% of all events) and 479,021 phone calls (6.3%). Due to the fact that the dataset covers millions of mobile phone interactions that have been recorded by a special application installed on students' phones, some issues discussed below arose during the preliminary data analysis phase.

Since events involving each student possessing the mobile phone were recorded independently from other mobile phones, some inconsistencies arose among which the most important was that not in all cases, the event was recorded on both sides. Since usually a couple of seconds pass between sending text message and receiving it by the recipients, we identified the same message by its length. However, sometimes, the recorded lengths were different, so in such cases, we were forced to make somewhat arbitrary decision based on the length difference whether the message was the same or separate one for which the corresponding event on the other side of communication was not recorded.

Next, when the students phoned each other and the call was not answered, the recipient's voice mail was reached, if it was enabled. Such case has been recorded as a regular phone call even if it has not been actually answered by the recipient. In order to solve this problem, we had to look for orphan records that do not have their corresponding records on the recipient's side.

This part of data preprocessing removed 12,482 records from our dataset. Afterwards, we performed an exploratory analysis of events. In Figure 5, we show the basic statistics regarding the communication patterns of the participants of the study.

The second part of the dataset includes surveys containing peers enumerated by the participants at the end of each term in response to the following question: "*In the spaces below, please list up to 20 people (friends, family members, acquaintances, or other people) with whom you spend time communicating or interacting*." To be more specific on how the process of obtaining data looked like, the students have been presented with a form containing twenty empty fields with the request in question—they had to fill the names by themselves without any suggestions or generators. As already mentioned in the main part of the manuscript, the ordering of the enumerated peers was not imposed, but even if the students would start with listing the most important and not necessary most frequent contacts, we could not assume that.

Lastly, the students were asked to enumerate in any order up to twenty peers with whom they interacted recently. However, the system recorded each name with the accompanying number representing name's position on the list. For some surveys, there were some holes in this numbering, so either the student did not list the peers accordingly (left some blank lines in between names) or partial results were not recorded. This is why we consider listed peers as the sets without any implied ordering, and the Jaccard sets comparison metric was used for evaluation purposes instead of any other that are based on ordering, e.g., Pearson or Spearman rank correlation coefficients.

## Data Availability

The code for computing the state of the social network built using the CogSNet model has been made public. A fast C implementation prepared by the authors of this manuscript is available as a Code Ocean capsule (https://doi.org/10.24433/CO.6088785.v1). The NetSense dataset used for the model validation is available for other researchers; the readers interested in accessing the data should contact the coauthor Professor Omar Lizardo.

## Conflicts of Interest

The authors declare that there are no conflicts of interest.

## Authors' Contributions

All authors participated in the design of the research and computational experiments. RM and MK ran the computational experiments and collected the results. All authors participated in analysis of the results. All authors wrote and edited the manuscript.

## Acknowledgments

The authors would like to thank Prof. Tomasz Kajdanowicz for his valuable remarks regarding statistical tests. This work was partially supported by the National Science Centre, Poland, by the following projects: 2015/17/D/ST6/04046 (RM) and 2016/21/B/ST6/01463 (PK, MK), Office of Naval Research (ONR) (Grant no. N00014-15-1–2640), Army Research Office (ARO) (Grant no. W911NF-16-1–0524), European Union's Marie Skłodowska-Curie Program (Grant no. 691152), and Polish Ministry of Science and Higher Education (Grant no. 3628/H2020/2016/2). Calculations have been carried out using resources provided by Wrocław Centre for Networking and Supercomputing (https://wcss.pl) (Grant no. 177). Collection of the NetSense data was supported by the U.S. National Science Foundation (Grant no. IIS-0968529). Further support for data collection was provided through University of Notre Dame and Sprint.